\documentclass[10pt,journal,compsoc]{IEEEtran}

\ifCLASSOPTIONcompsoc
  \usepackage[nocompress]{cite}
\else
  \usepackage{cite}
\fi

\ifCLASSINFOpdf

\else
\fi

\usepackage[T1]{fontenc}
\usepackage{txfonts}
\usepackage[pdftex]{hyperref}
\usepackage{color}
\usepackage{siunitx}
\usepackage{booktabs,arydshln}
\usepackage{textcomp}
\usepackage{multirow}
\usepackage[caption=false]{subfig}
\usepackage{microtype} 
\usepackage{ccicons}  
\usepackage{mathptmx}
\usepackage{graphicx}
\usepackage{times}
\usepackage{enumitem}
\usepackage{tabu}
\usepackage{float}

\usepackage{cite}

\makeatletter
\def\adl@drawiv#1#2#3{%
        \hskip.5\tabcolsep
        \xleaders#3{#2.5\@tempdimb #1{1}#2.5\@tempdimb}%
                #2\z@ plus1fil minus1fil\relax
        \hskip.5\tabcolsep}
\newcommand{\cdashlinelr}[1]{%
  \noalign{\vskip\aboverulesep
           \global\let\@dashdrawstore\adl@draw
           \global\let\adl@draw\adl@drawiv}
  \cdashline{#1}
  \noalign{\global\let\adl@draw\@dashdrawstore
           \vskip\belowrulesep}}
\makeatother

\begin{document}

\title{
Geono-Cluster: Interactive Visual Cluster Analysis for Biologists
}

\author{Bahador Saket$^{*}$\thanks{Bahador Saket and Subhajit Das contributed equally to this work.}, Subhajit Das$^{*}$, Bum Chul Kwon, and Alex Endert

\IEEEcompsocitemizethanks{\IEEEcompsocthanksitem Bahador Saket, Subhajit Das, and Alex Endert are with Georgia Institute of Technology. E-mail: 
 {saket, das, endert@gatech.edu}
\IEEEcompsocthanksitem Bum Chul Kwon is with IBM Research. E-mail: bumchul.kwon@us.ibm.com.
}
}

\IEEEtitleabstractindextext{%
\begin{abstract}
Biologists often perform clustering analysis to derive meaningful patterns, relationships, and structures from data instances and attributes. Though clustering plays a pivotal role in biologists' data exploration, it takes non-trivial efforts for biologists to find the best grouping in their data using existing tools. Visual cluster analysis is currently performed either programmatically or through menus and dialogues in many tools, which require parameter adjustments over several steps of trial-and-error. In this paper, we introduce Geono-Cluster, a novel visual analysis tool designed to support cluster analysis for biologists who do not have formal data science training. Geono-Cluster enables biologists to apply their domain expertise into clustering results by visually demonstrating how their expected clustering outputs should look like with a small sample of data instances. The system then predicts users' intentions and generates potential clustering results. 
Our study follows the design study protocol to derive biologists' tasks and requirements, design the system, and evaluate the system with experts on their own dataset.
Results of our study with six biologists provide initial evidence that Geono-Cluster enables biologists to create, refine, and evaluate clustering results to effectively analyze their data and gain data-driven insights.
At the end, we discuss lessons learned and implications of our study.
\end{abstract}

}

\maketitle

\IEEEdisplaynontitleabstractindextext

%
\IEEEpeerreviewmaketitle

\ifCLASSOPTIONcompsoc

\IEEEraisesectionheading{\section{Introduction}\label{sec:introduction}}
\else
\section{Introduction}
\label{sec:introduction}
\fi

\IEEEPARstart{C}{lustering} is the task of summarizing and aggregating complex multi-dimensional data in such a way that items in the same group are more similar to each other than those in different groups. 
Domain experts often want to perform clustering to find groups of data items that share common characteristics with respect to data attributes. For example, a biologist who wants to investigate genome data can cluster gene sequential data according to similarity between their expression profiles. Since clustering does not require hand-annotated labels for supervision, it is generally used as a flexible method for exploratory data analysis. As such, clustering has a widespread application in several domains including biology~\cite{nugent2010overview}, chemistry~\cite{downs2002clustering}, and social sciences~\cite{bartholomew2008analysis}.

Our paper aims to accommodate the process of interactive visual clustering for biologists. Like other domain experts, biologists also want to cluster their data and visualize the result to investigate patterns, relationships, and structures among data instances and attributes. However, not all biologists often have formal data science training. The lack of knowledge in data science often prevents users from clustering their data and from interpreting the results in the biological context using the existing tools. Based on our collaborations with a group of biologists, we found that they use tools such as SAS and/or programming languages like R to run cluster analysis on their data. These tools require users to specify clustering algorithms and parameters in written scripts. The absence of user-friendly tools may increase execution costs and impede the adoption of clustering methods for data exploration.

There is a large body of visual analytic systems that employ visual clustering as a part of high dimensional data analysis (e.g.,~\cite{desJardins:2007, Clusterophile2, VISTA, 1016905, DICON, Jeong:2009, clusterVision}). 
Some of these visual analytic systems are often complex, and require careful tuning, steering, and parameterization of the clustering models. 
Interaction complexity in such systems often poses fundamental usability challenges for those domain experts who may not have formal data science training~\cite{BeyondControlPanel}. 
Furthermore, it is challenging for domain experts to directly apply their knowledge into the clustering processes. 
For example, biologists exploring genome data might want to merge two clusters because of the similarity of evolutionary history of the genes located in two clusters. Alternatively, they might want to subdivide a specific cluster to estimate the disease risk of genes in different sub-clusters in a specific population. As such, current tools are ill-equipped to help biologists build groupings based on their prior knowledge and to discover patterns from the results.
However, many of the existing tools do not provide visual guidance on how to reach desirable results or translate users' analytic goals into a proper setting of algorithm and parameter.


To tackle the challenges for biologists, we present \textbf{Geono-Cluster}, a visualization tool that applies the ``by demonstration''~\cite{CGASaket, Saket:visbyDemo} paradigm. 
Instead of requiring biologists to transform their clustering tasks into system specifications by going through layers of menus or programming it, Geono-Cluster allows biologists to directly apply their domain expertise by visually demonstrating how their expected changes should look like (e.g., dragging one cluster and dropping it over another cluster to show their interest in merging the clusters). 
By translating these demonstrations into numerical processes that update the underlying cluster distance functions, the system predicts biologists' intentions and generates potential clustering results (e.g., different visual clustering outputs that merged those two clusters). 
We have developed Geono-Cluster in collaboration with biologists investigating disease risks frequency across different populations. 
We closely followed the design study protocol~\cite{sedlmair2012design} to derive system requirements, tasks to be supported, and design guidelines based on feedback from biologists.

To evaluate our approach, we then conducted a qualitative study with six expert biologists at the Georgia Tech. In this evaluation, we observed how our tool helps biologists to cluster their data and identify challenges they encounter while using our tool. 
We also conducted a semi-structured interview to collect biologists' feedback and new ideas.
Our results demonstrate that Geono-Cluster enables biologists to build, refine, and evaluate clustering outcomes with intuitive demonstration-based interaction and to interactively explore the results through multiple views.

\section{Related Work}
In this section, we explain the importance of clustering for exploratory data analysis. 
We also review interactive tools and methods designed for clustering data, and describe the challenges that domain experts often encounter while working with these tools. 


\subsection{Clustering}
Different domains are seeing a surge in data collection at an alarming rate, which needs to be efficiently analyzed~\cite{stark2006biogrid}. Clustering a dataset is a proven approach to understand the inherent structure of large datasets and is used in several critical domains~\cite{nugent2010overview,bartholomew2008analysis, downs2002clustering}. There exist a plethora of tools and programming languages that support cluster analysis, such as R\cite{R}, Matlab \cite{Hunter:2007}, SAS~\cite{SAS}, and Python\cite{Python}. Users can choose from a variety of different clustering algorithms and their hyperparameters depending on their analysis goals. Despite the flexibility, using such methods often require an intermediate to advanced knowledge of programming skills. For this reason, domain experts often need to go through a steep curve of learning these programming languages. Furthermore, these tools often lack visual feedback and interactivity, which make users difficult to understand the results and to reconfigure the setting for improved results in next iterations.
Thus, the lack of interactivity can increase execution costs and impede the data exploration process~\cite{card:1991:information}. 




\subsection{Interactive Visual Clustering Analysis}
Researchers have been investigating various techniques and approaches to facilitate interaction in clustering analysis, with the goal of bringing a human in the loop. Effective user interaction is critical to the exploratory data analysis process, and thus to the success of the visual analytic systems for visual clustering. A large body of previous work designed and implemented interactive tools to support interactive visual clustering and analysis (e.g.,~\cite{clusterVision, Wenskovitch2017, ClustroPhile, SomeFlow, Basu:2010, Clusterophile2, ClusterSculptor,Hu:Clustering, Liu2012, Guo:Clustering, Andromeda, DubeyClusterLevel}). Here we discuss some of the most related tools to our work. 

Clusterophile~\cite{ClustroPhile} and Clusterophile 2~\cite{Clusterophile2} are both designed to enable users to explore different choices of clustering parameters and reason about clustering instances in relation to data dimensions. 
Datta et al. built an interactive clustering system - CommunityDiff, showing a mechanism to visualize ensemble space by using a weighted combination of various clustering algorithms to aid identifying patterns, commonalities, and differences \cite{CommunityDiff}.
iVisClustering~\cite{iVisClustering} is another tool that supports document clustering based on a widely used topic modeling method called latent Dirichlet allocation (LDA). Hu et al.~\cite{Hu:Clustering} and Guo~\cite{Guo:Clustering} separately developed interactive tools that allow users to select features while clustering their data. ClusterSculptor~\cite{ClusterSculptor} is another tool that aids data scientists in the derivation of classification hierarchies in cluster analysis. VisBricks~\cite{VisBricks} provides multiform visualization for the data represented by clusters (it enables users to select which visualization technique to use for which cluster). In a different project, Basu et al.~\cite{Basu:2010} propose a tool that allows users to move data items and build clusters of data items from a larger set, while the system suggests data items which can be further added to the set. ClusterVision~\cite{clusterVision} is a more recent tool that enables users to cluster data using a variety of clustering techniques and parameters and then ranks clustering results utilizing five quality metrics.
It also allows users to interpret the clustering results using multiple, coordinated views and to interactively filter results by setting various constraints on data instances.

Geono-Cluster differentiates itself from the aforementioned work mainly by supporting biologists' visual clustering analysis. The existing visual analytic systems often require careful tuning, steering, and parameterization of the clustering models~\cite{iVisClustering,ClusterSculptor, Clusterophile2}. In such systems, analysts need to translate their analytic goals into clustering specifications by going through layers of menus. Unlike existing tools, Geono-Cluster enhances user interaction expressivity by enabling users to interactively define clustering results by their demonstration on data items, which is more user-friendly and easy to understand for domain experts.

\subsection{Interactive Clustering Analysis for Biologists}
There exist some visualization tools that are designed for clustering analysis of biological data.
StratomeX~\cite{Lex:Stratomex} is an interactive visualization tool that enables users to explore the relationships of subtypes across multiple genomic data types. StratomeX is mainly designed to support tasks with ``comparative nature'' (e.g., evaluate how well two or more stratifications support each other). CComViz~\cite{ccomviz} is a different application that uses the parallel sets technique to compare clustering results. Kern et al. proposed novel methods for evaluating and comparing cluster results and implemented their methods into StratomeX~\cite{2017_bmc_clustering}. XcluSim~\cite{XclusSim} is another tool for bioinformatics data helping users to compare multiple clustering results, supporting a diverse set of algorithms. XcluSim combines several small sub-views to form a multiview layout for cluster evaluation.

Unlike other tools that are mainly designed to support tasks with ``comparative nature'', Geono-Cluster is designed to cover a different category of tasks such as creating customized clusters, and merging and splitting clusters. Moreover, Geono-Cluster aims to reduce biologists' cognitive cost and enhance interaction expressivity by implementing the ``by demonstration'' approach. Geono-Cluster enables biologists to apply their domain expertise into clustering processes by interacting with visuals representing data items and clusters (e.g., a biologist can express that a data item does not belong to a cluster by dragging it out of the cluster). As a result, the system finds the most appropriate clustering results based on the user interactions (e.g., finding clustering results where the selected data item does not belong to the specific cluster).

\subsection{The Demonstration-Based Interaction }
The demonstration-based interaction has been applied to many applications. A common application of the technique in human-computer interaction is ``programming by
demonstration.'' Cypher states in his article~\cite[page 1]{cypher1993watch}: \textit{``The motivation behind Programming by Demonstration is simple and compelling: if a user knows how to perform a task on the computer, that should be sufficient to create a program to perform the task. The user should be able to instruct the computer to ``Watch what I do'', and the computer should create the program that corresponds to the user's actions.''} 

Other domains that have successfully used the ``demonstration-based'' paradigm include data cleaning~\cite{Lin:dataCleaning, Kandel_wrangler}, database querying~\cite{Zloof1975QueryBE}, temporal navigation~\cite{kondo2014dimpvis}, visual data analysis~\cite{EMS}, and visualization construction~\cite{Saket:visbyDemo,CGASaket}. For example, Kandel et al.~\cite{Kandel_wrangler} enables users to demonstrate desired changes to tabular dataset by making direct edits to the table elements (e.g., select and delete empty rows)~\cite{Saket2019InvestigatingDM}. In response to the given demonstrations, the system suggests potential trasnformations that may be applied to generalize the demonstrated change and update the data table. Previous work also enabled users to steer the dimension reduction models by demonstrating the relative similarity between data items (e.g., ~\cite{disfunction, endert2012semantic, kwon2016axisketcher}. 

Our work enables biologists to demonstrate their desired clustering results by directly manipulating visual elements representing clusters (e.g., moving a subset of data items from one cluster to another). In response to users' demonstrations, the system computes possible clustering results and recommends them. Inspired by previous work~\cite{Saket:visbyDemo,Kandel_wrangler}, each recommendation provides a visualization which gives an overview of the clustering result and a textual explanation.

\section{Formative Assessment}
In this section, we explain a formative assessment that we conducted to characterize our domain experts' workflow, derive tasks and requirements from it, and generate design guidelines to design Geono-Cluster.

\subsection{Characterizing domain experts, data, and tasks}
The motivation of this work stems from an ongoing project in which we have been collaborating with biologists at the Georgia Tech. We have been working with the biologists over the past 13 months to design and build solutions for supporting interactive visual clustering of disease risk factors. 

The dataset used by the biologists is from Genome Wide Association Studies (GWAS Catalog)~\cite{GWAS_catalog} which includes published SNPs (single-nucleotide polymorphisms, 
representing differences in a single DNA building block, called a nucleotide),
and association studies to analyse genetic sequences.
Through this dataset, biologists intend to determine ``alleles'' that correlate to various diseases and traits. 
Alleles are various forms of a gene that are formed by mutation and are found at the same place on a chromosome.
Using GWAS dataset biologists analyse SNPs to find how do they vary between various genome samples. 


During data analysis, biologists often focus on certain features of their dataset such as, disease/trait, SNP identification number, risk allele frequency, p-value, and odds ratio/beta. Focusing on those values, they try to answer questions like, how and why disease risk frequencies differ across populations, what are the statistical power to detect those known SNPs, and how well associations found in one population can transfer/replicate well to another population. To answer such questions, researchers cluster their data to investigate patterns and relationships of position on the genome, risk allele, and risk allele frequencies that impact diseases risk frequencies across different populations. This is an iterative process and biologists frequently create customized clusters, merge/split clusters, and investigate sub-clusters within a specific cluster to test their hypotheses based on their expertise. 

To cluster and visualize their data, these biologists currently use tools/programming languages like R~\cite{R} and SAS~\cite{SAS}. They revealed that the current process of clustering and visualizing the data is rather time-consuming, cumbersome, and occasionally error-prone. One of the biologists stated that: \textit{``my process [data exploration process] is sometimes slow. [...] I search for code snippets online. After finding the code, it takes 2 or 3 trials to get the code working.''} 
Our observations as well as researchers' feedback show that they often need to write and execute scripts. 
Writing scripts becomes even more challenging when they want to perform more specific tasks such as merging two or more clusters. 
Moreover, the interactive visual clustering tools are ill-equipped to support specific and customized tasks often performed by these biologists. To overcome the challenges, we aimed to design an interactive visual clustering tool that enables intuitive and fast visual data clustering for biologists.



   \begin{figure*}[t]
        \includegraphics[width=1\linewidth]{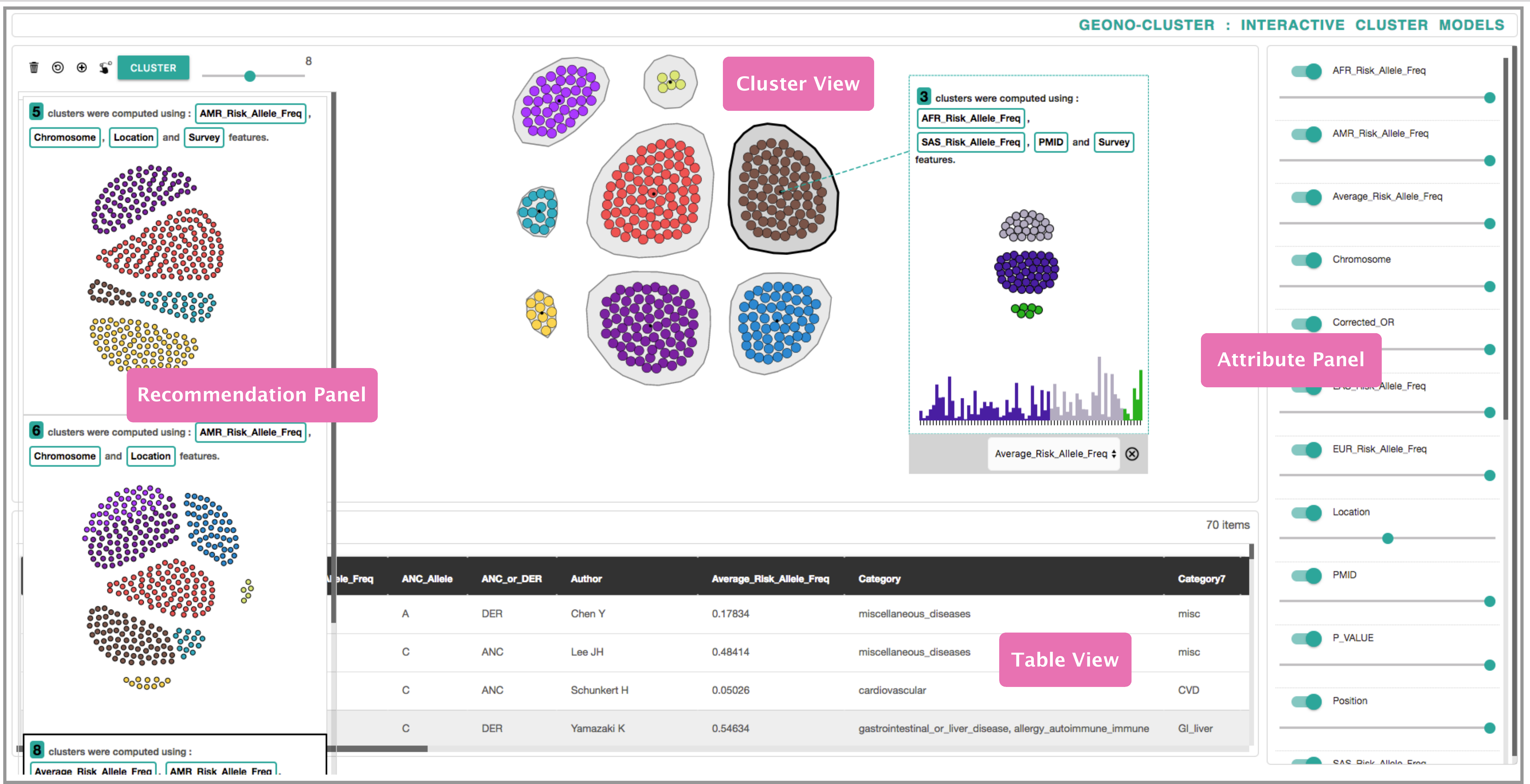}\vspace{-1.6em}
        \caption{The Geono-Cluster user interface consists of a Cluster View, a Recommendation Panel, a Table View, and an Attribute Panel. Cluster view visualizes the clustered data and provide a medium for users to provide visual demonstrations. Recommendation panel shows different clustering results based on the demonstrations provided by users. Table view shows a tabular representation of the loaded dataset. Attribute Panel lists the attributes of the loaded dataset and their weights.}
        \label{fig:Geono-Cluster}
        \vspace{-0.3cm}
    \end{figure*}

\subsection{Tasks and Requirements} \label{sec:tasks}
Following a user-centered method~\cite{norman1993things}, we began our iterative design process by investigating current practices, needs, and challenges. We conducted multiple group discussions with two biologists at the Georgia Tech. We started our discussions with the biologists by asking them: 1) what kinds of questions do they ask and answer while exploring their data? 2) why do they perform clustering tasks during their analysis?, and 3) how do they currently create clusters? Then, we freely continued our conversation that touched upon the tools, analytic methods, and challenges they face during the process. We took notes during all the group discussions. We then read through our notes to gain a better understanding of the requirements and challenges these biologists encounter while clustering their data. After reading the data, we identified the meaningful text segments (e.g., \textit{``[...] here we combine these two clusters.''}). We then assigned a code phrase that describes the meaning of the text segment (e.g., merging clusters). We initially identified three commonly performed clustering operations that are currently challenging for biologists to complete using existing programming languages and tools.\vspace{0.5em}

\noindent\textbf{T1: Hand-craft, Merge, and Split Clusters:} Biologists apply their domain knowledge to create customized clusters to better understand which factor(s) is causing the ascertainment bias on the dataset that are being used popularly. For example, one of the biologist stated: \textit{``Given the identified SNPs [single-nucleotide polymorphisms] that are associated with common disease and traits, it's interesting to create a cluster of SNPs.''} In addition, biologists apply their domain expertise to merge or split two or more clusters depending on how related they think the clusters are based on given feature(s). For example, one of the biologists mentioned: \textit{``Depending on the evolutionary history of the genes, two or more clusters can be really related to each other. If ascertained they are related, we will merge them as one cluster."} Another biologist reported that \textit{``In my new project, we are comparing Africans to non-Africans. In this case I merge Americans, East Asians, and Europeans as one cluster, and compare that to Africans data.''} \vspace{0.5em}
    
    \noindent\textbf{T2: Divide each cluster to sub-clusters:} Biologists often investigate sub-clusters within a specific cluster to: 1) understand which other factors can affect the cluster, 2) compare two clusters based on the member data items in each, 3) see trends and patterns in the sub-clusters, with respect to chosen features, and 3) further investigate the risk of disease across a subset of data within a cluster. We noticed that the biologists found existing solutions challenging because they had to write lines of scripts to compute and visualize sub-clusters in a given cluster. Furthermore, the existing methods prohibit rapid iteration and visualization of results, which inevitably prolongs the exploratory cluster model construction process to understand their data better.\vspace{0.5em}
    
    \noindent \textbf{T3: Adjust feature contributions:} Biologists need to easily see by how much different attributes/features contribute to computing a cluster. Moreover, they often need to adjust the importance of different features used for computing a cluster. Biologists currently have to programmatically adjust the importance of features, execute the code, and visualize the outcome. They often repeat this process multiple times until they achieve a satisfactory result. They need interactive methods to view and refine feature contributions.


\subsection{Design Guidelines}
We needed to explore alternatives and make design decisions to better support the aforementioned tasks.
In particular, Geono-Cluster should be easy to use by experts who do not have formal data science training. 
We developed a set of design guidelines to inform those interested in developing visual analytic tools for domain experts (in particular biologists). These guidelines are based on existing tools designed for supporting visual data exploration for biologists~\cite{XclusSim, 2017_bmc_clustering}, mixed-initiative systems~\cite{Horvitz:1999}, and our experiences through several design iterations with biologists.\vspace{0.5em}

\noindent \textbf{G1: Shifting the burden of specification from the biologists to the systems.} The existing tools and technologies put the burden of specification on biologists. For instance, while clustering their data, biologists need to specify the clustering algorithm, number of clusters and iterations, and other parameters. Biologists find the current process time-consuming and cumbersome. Instead of requiring biologists to specify the clustering models by programming or going through layers of menus, the tool should provide an environment that enables them to visually demonstrate how the expected clustering outcomes should look like. By translating the given demonstrations, the system could estimate the biologist's intention and generate appropriate results. This way we could balance the responsibility between the biologist and the system -- biologists provide visual demonstrations, based on this, the system infers potential clustering results and recommends them. 

\vspace{0.5em}

\noindent \textbf{G2: Allow biologists interaction to drive recommendations.} As biologists explore their data, their interests will evolve. As a result of this, biologists need to explore various clustering models rapidly. To support this need, the system should recommend potential cluster models that biologists should consider. Furthermore, the clustering recommendations should be adapted for biologists' analytic goals at the time. The recommendation engine should steer multiple clustering models based on biologist-specified expected visual outcomes. In addition, biologists can also directly adjust feature contributions to update the clustering results.
In aggregate, these interactions create demonstrations which serve as the primary units by which biologists communicate their expected changes to the system. \vspace{0.5em}

\noindent \textbf{G3: Enhance interpretability of recommendations.} Biologists reported their interest in seeing more details about different clustering results while skimming through different recommendations. However, not all biologists might be familiar with technical terms used to describe a cluster such as silhouette value. Therefore, recommended clustering results should be presented in a transparent manner so that biologists can extract the most important and understandable information (e.g., contributing features) used for clustering results.

\section{Geono-Cluster}
Based on the tasks and guidelines, we developed Geono-Cluster, a visual clustering tool for biologists. 
All components of the Geono-Cluster were implemented using JavaScript and Python. 
The visualization modules are built using the D3.js toolkit.
 
 \subsection{Usage Scenario}

In this section, we motivate the design of our system and illustrate
the functionality via a usage scenario. We indicate how a domain expert can
utilize Geono-Cluster to perform visual cluster analysis on the GWAS Catalog dataset~\cite{GWAS_catalog}. This dataset includes detailed information regarding the identified single-nucleotide polymorphisms (SNPs) associated with common diseases and traits (e.g., position on the genome, risk allele frequencies, p-value, effect sizes, etc.). 
SNP is a region on the gene where more than one allele (A, C, G, T) is observed and each row on the dataset is a SNP \cite{guidetogenomedata}.







Megan is a biologist who wants to compare populations from the \textit{GWAS} dataset to understand disease risk factors related to geographical regions (e.g., if gene samples collected from ``America'' are more prone to cancer than gene samples collected from ``Europe''). She launches Geono-Cluster to cluster the data, and to compare associated sub-populations. First, Megan skims through different features on the Table View (see Figure \ref{fig:Geono-Cluster}).  

\begin{figure}[tb]
        \includegraphics[width=3.4in]{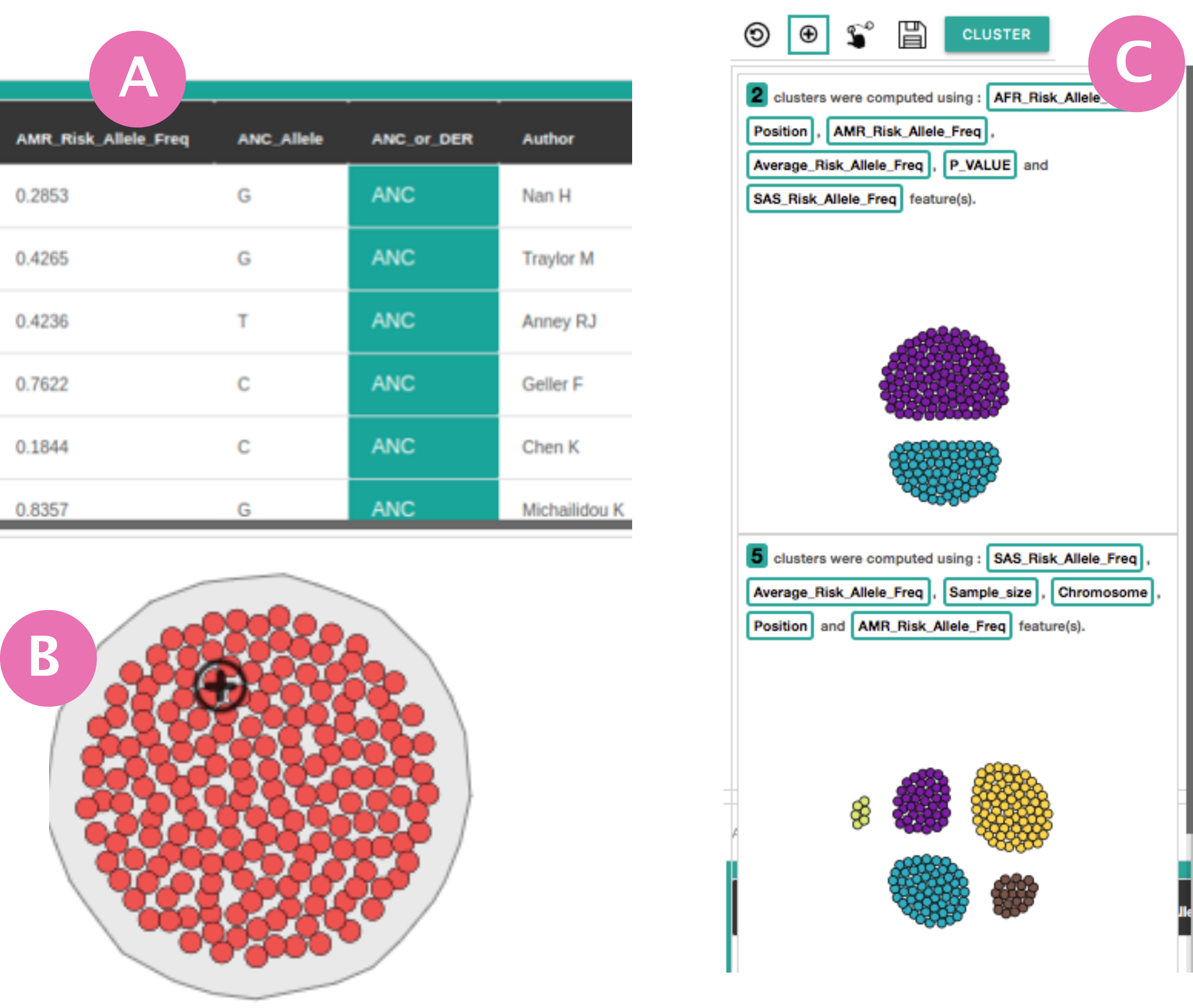}
        \caption{ \textbf{A)} Megan clicks on a cell in the column \textit{ANC-or-DER} with the value \textit{ANC}. The system automatically selects all data items with ancestory \textit{ANC}. \textbf{B)} She drags the selected data items and drops them to the cluster view. The system automatically represents data items as red circles and places them in an independent cluster. \textbf{C)} The system also recommends potential clustering layouts based on the demonstration provided by Megan.}
        \label{fig:us_drag} 
        \vspace{-0.5em}
\end{figure}

Megan knows that there are two types of gene samples: ANC and DER. 
ANC samples are the genes that are derived from either humans or monkeys. 
DER are the gene samples that are derived from the mixture of humans and monkeys. 
Megan starts her exploration by comparing the disease risk factor between the two types of genes samples based on their ancestry. 
Megan first skims through different features to find the \textit{ANC-or-DER} feature on the Table View. 
She clicks on a cell in the column \textit{ANC-or-DER} with the value \textit{ANC} in the Table View to demonstrate her interest in selecting all the data items with ancestry \textit{ANC}. 
In response, Geono-Cluster automatically selects all data items with ancestry \textit{ANC} (see Figure~\ref{fig:us_drag}-A). Megan then demonstrates her interest in clustering data items with \textit{ANC} value by dragging them from the Table View and dropping them to the Cluster View. In response to the demonstration, the system automatically represents data items as red circles and places them in the \textit{Red Cluster} (see Figure~\ref{fig:us_drag}-B). At this point, the system also recommends potential clustering results based on the demonstration provided by Megan (see Figure~\ref{fig:us_drag}-C).
Even though Megan's interactions may lead to grouping the data based on the chosen categorical data attribute (\textit{ANC or DER}), in essence, this is a start to allow a user demonstrate their intent to find a clustering model that represents agreeable clusters in the data. 
Their interactions are inferred as implicit intents by the system to find the most appropriate cluster model as opposed to just group the data by a set of categorical variables.

\begin{figure}[tb]
 \centering 
\includegraphics[width=\columnwidth]{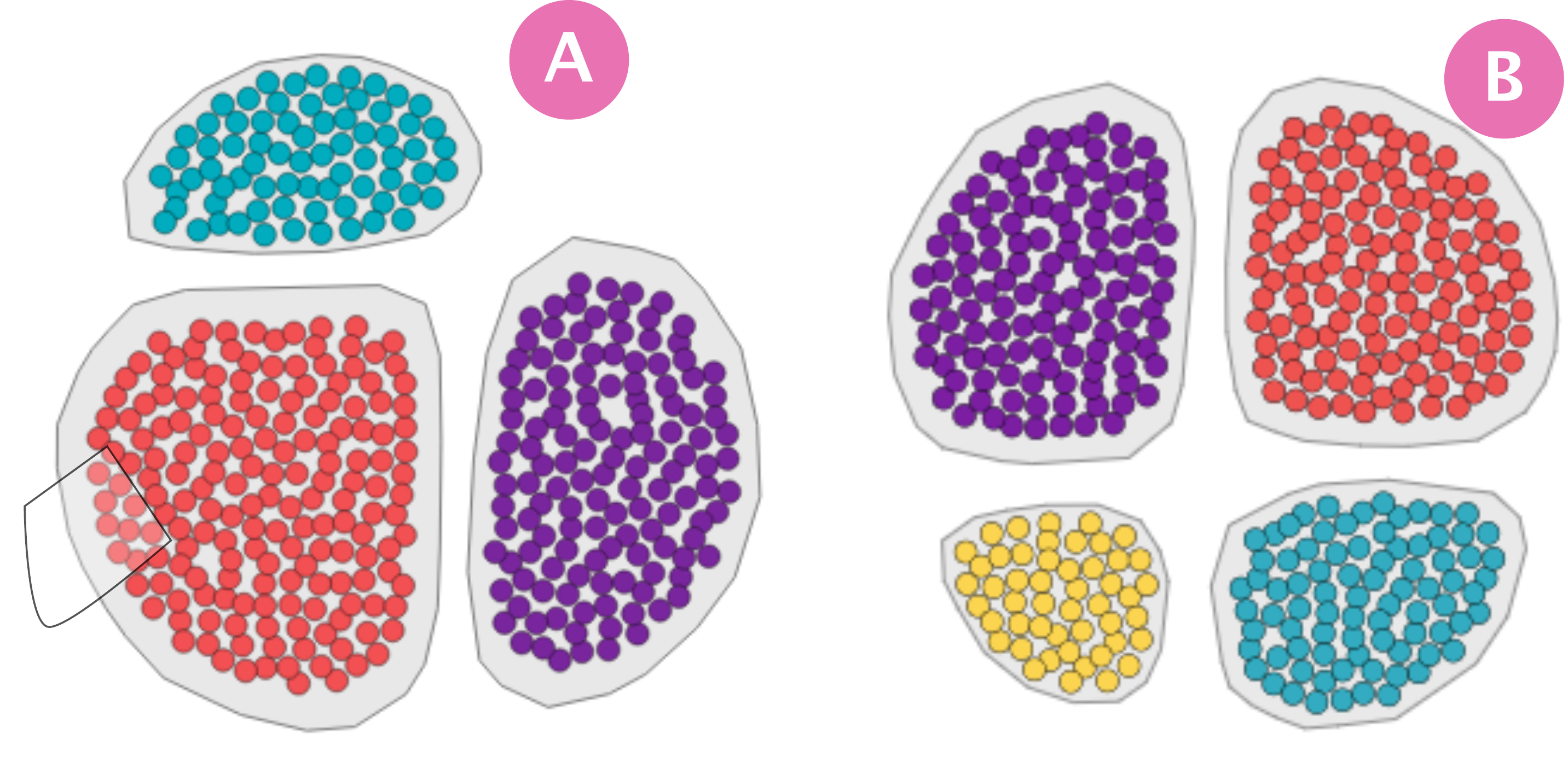}
        \caption{\textbf{A)} Megan uses lasso tool to select a subset of data items from Cluster $1$. 
        \textbf{B} The system automatically finds other similar data items and defines cluster 4 containing all these data items. }
        \label{fig:us_recom_exist}\vspace{-0.5em}
\end{figure}




    
Megan opens the recommendation panel and previews other clustering options through the thumbnail previews. She finds one of the clustering results recommended by the system interesting. She clicks on this thumbnail, which updates the Cluster View with the recommended cluster layout by adding the \textit{Blue Cluster} and the \textit{Purple Cluster} (along with the \textit{Red Cluster}) in the Cluster View (see Figure~\ref{fig:us_recom_exist}-A).   

Megan explores the data items within each cluster by hovering over each data item to see its details. After exploring a few data items, she notices that while the  \textit{Red Cluster} contains genome samples with ancestry \textit{ANC}, the recently added clusters (\textit{Blue Cluster} and the \textit{Purple Cluster}) contains all the gene samples with ancestry \textit{DER}. She further notices that most of the items in the \textit{Blue Cluster} have the chromosome value higher than $8$, and the genome samples belong to the region \textit{America}. Now she understands what each of the three clusters represent.

Next, Megan demonstrates her interest of excluding data items with ancestry \textit{ANC} that belong to \textit{Africa} from the \textit{Red Cluster}. To do so, she lasso-selects a subset of data items with ancestry \textit{ANC} that belong to the region \textit{Africa} from the \textit{Red Cluster} (see Figure~\ref{fig:us_recom_exist}-A). 
Important to note that points closer to each other in a cluster are expected to be similar to one another based on the applied cluster model. Thus using the lasso selection, Megan selects similar points from a cluster for further analysis.
She then drag-and-drops these points out of the cluster. In response, the system automatically finds other similar data items with ancestry \textit{ANC} that belong to the region \textit{Africa}, and then defines the \textit{Yellow Cluster} containing these data items.Further, the system updates the recommendations in the recommendation panel accordingly.

\begin{figure}[tb]
 \centering 
\includegraphics[width=0.8\columnwidth]{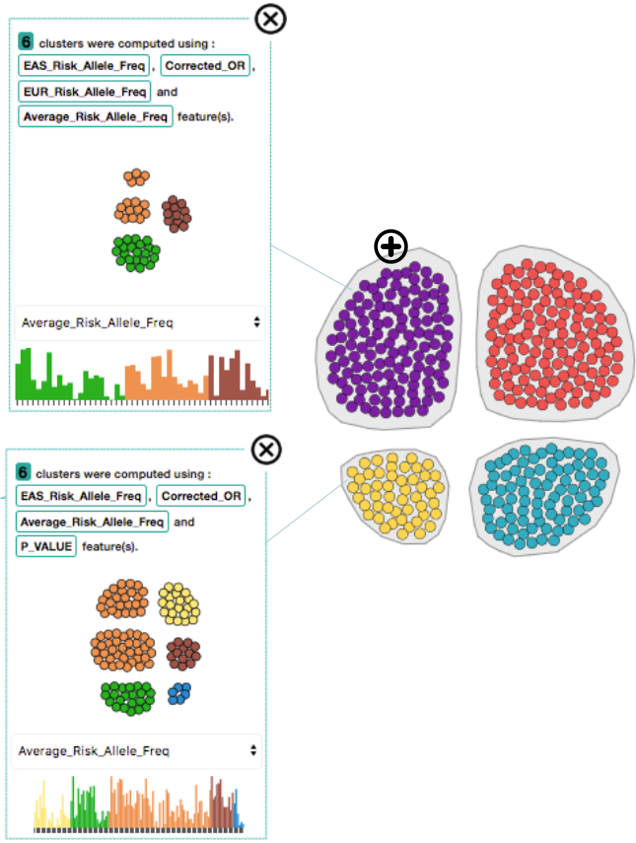}
        \caption{Megan clicks on the ''+'' icon to open the sub-cluster panel for clusters 1 and 4. Bar chart views showing comparisons of the feature Average-Risk-Allele between Cluster 1 and 4.}
        \label{fig:comparison} \vspace{-1em}
\end{figure}

Looking at the \textit{Purple Cluster}, and the \textit{Yellow Cluster}, Megan realizes that the items in these two cluster are with ancestry \textit{ANC} and \textit{DER} respectively.
Megan wants to compare the distribution of the feature \textit{Average-Risk-allele} between these two clusters to compare their disease risk factors. 
She clicks on the ''+'' icon (see Figure~\ref{fig:comparison}), which is shown upon hovering on a cluster, to open the sub-cluster panel for each cluster. 
Each sub-cluster further groups the data items per cluster.
In addition, the sub-cluster panel contains a bar chart view highlighting the distribution of a chosen feature (\textit{Average-Risk-allele}) through a drop-down selector for all data items in the parent cluster (see Figure~\ref{fig:comparison}).
After inspecting the distributions, Megan does not notice any significant difference in the value of \textit{Average-Risk-Allele} between the\textit{Purple Cluster} and the \textit{Yellow Cluster}.


Megan explores the \textit{Blue Cluster} (with \textit{DER} ancestory) to inspect its sub-cluster layout and distribution of the feature \textit{Average-Risk-Allele}.
When she compares the distribution of feature contributions of the \textit{Yellow Cluster}, she discovers that the genes sampled from \textit{Africa} with ancestory \textit{ANC} has much higher disease risk factor than those sampled from other regions with ancestry \textit{DER}. 

Megan decides to merge the clusters \textit{Blue} and \textit{Purple}. To do so, she demonstrates her interest in merging the clusters by dragging the \textit{Blue Cluster} and dropping it over the \textit{Purple Cluster}. In response, the system recommends new cluster layouts on the Recommendation Panel. 
She previews the thumbnails from the recommendation panel and selects the second recommendation, which results in placing $3$ clusters in the Cluster View. To continue discussing the findings and implications with other colleagues, she exports a \textit{.png} screenshot of the current cluster layout. 
She also saves the results as a~\textit{.csv} file to investigate them more in other programs like R and SPSS.

\subsection{Views and User Interface}
Geono-Cluster's interface consists of four views: a Cluster View, a Recommendation Panel, a Table View, and an Attribute Panel. See Figure~\ref{fig:Geono-Cluster} for more details.\vspace{0.5em}

\noindent\textbf{Cluster View} visualizes the clustered data as Figure~\ref{fig:Geono-Cluster} shows. For testing their hypotheses, biologists often perform actions at the level of data items (e.g., move data items from one cluster to another). We visually present each cluster and its members on the Cluster View. The colored circles in each group represent members of a cluster; the surrounding hull represents the cluster. Users can hover over a circle, which prompts relevant attribute details of the data. Users can specify the number of clusters using the slider shown on the top-left. Cluster View is an environment similar to a spatial workspace in which users can move data items to structure their information and provide visual demonstrations \textbf{(G1)}. For example, a biologist might notice a set of data items should not be in a specific cluster. Thus, she can demonstrate that those points belong to a different cluster by dragging them from one cluster to another. The system uses the visual demonstrations provided by the users to steer the underlying recommendation engine \textbf{(G2)}.\vspace{0.5em}

\noindent\textbf{Recommendation Panel} shows different clustering results. Based on users' demonstrations on the Cluster View, the system recommends a set of appropriate clustering outputs. To compute the recommended clustering results, the underlying recommendation engine takes into account different (1) clustering techniques/algorithms; (2) combinations of attributes/features; and (3) clustering hyperparameters (i.e., varying 'k' for k-means clustering technique). Read section \ref{Sec:comp_tech} for more details.

During the design process of Geono-Cluster, we examined different ways of presenting recommended clusters.
We first considered showing all the recommended clustering results as small thumbnails in the Recommendation Panel. 
The biologists liked the idea and the way that we recommended clustering results. 
However, the main challenge that biologists encountered was that they were not able to infer detailed information from the small thumbnails. 
Thus, they requested adding textual description of details about each clustering result in the recommendation. 
Currently, each thumbnail includes a textual description about the number of clusters, features used to compute the clustering recommendations, and a visualization of the clustering result showing distribution of data items over clusters \textbf{(G3)}. 

Initially, we designed the recommendation module to update the view with new clustering recommendations whenever users show their demonstrations and/or adjust feature contributions. However, our users revealed that such approaches may distract their ongoing investigations on the current results. Thus, we compute cluster recommendations in the background but do not show the results immediately. Once the computation is done, a notification pops up, encouraging users to explore the results on demand by toggling the `show recommendations' button (see Figure~\ref{fig:Geono-Cluster}). 
\vspace{0.5em}

\noindent\textbf{Table View} shows a tabular representation of the loaded dataset where each row is a data item (see Figure~\ref{fig:Geono-Cluster}). The initial version of Geono-Cluster did not include the Table View. 
However, biologists requested adding this view since it enabled them to check the raw data.  
The table updates on user's interaction on Cluster View. 
For example, selecting the hull of a specific cluster updates the Table View to show the details of items in the selected cluster. 
Users can click on a cell to find similar rows whose value are similar to the value of the item in that cell. This operation works on both quantitative and categorical data types. For categorical and ordinal data attributes we consider two data items similar if there is an exact match between them. For the numerical variables we define a threshold to measure similarity between two values.
This technique allows users to filter and select a subset of data instances (enables selection of multiple rows simultaneously).
\vspace{0.5em}


\noindent\textbf{Attribute Panel} lists the attributes of the loaded data set as Figure~\ref{fig:Geono-Cluster} shows. Users can turn on and off a set of attributes which directly affects the clustering algorithm. 
Furthermore, users can also adjust attribute contributions, specifying relative importance of the selected attributes to define cluster memberships \textbf{(G2)}.


\subsection{Interactions}
In this section we discuss how Geono-Cluster supports interactive operations commonly performed by biologists.  \vspace{0.5em}

\noindent\textbf{Merging and Splitting Clusters (T1)}: 
To merge two or more clusters, users first click on a cluster. They then demonstrate their interest in merging two clusters by drag-and-dropping the cluster on top of another cluster. Users can drag point(s) out of the cluster and drop into either i) another cluster or ii) a blank space (on the Cluster View). Drag-and-drop items into blank space is translated as forming a new cluster of the selected items outside the current cluster (see Figure~\ref{fig:us_drag}). Demonstration-based cluster customization enables users to interact with the data directly and removes any mid-level instruments such as control panels or menus. 

The merge interaction is derived from the previous work by Sarvghad et al.~\cite{EMS}, in which they enabled HIV researchers to merge bars in bar charts by dragging one bar and dropping it over another bar. 
Biologists liked this interaction design and found it ``direct and intuitive''. 
To split clusters, we initially enabled biologists to select the data items by clicking on each circle representing a data item. 
However, biologists found it cumbersome and time-consuming. So, we implemented the lasso-selection such that users can select multiple data items easily. This operation allows user to brush over a set of data samples (represented as circles) in the Cluster view. In response the system extracts those samples from the current cluster and places them in a new cluster. If data samples from multiple clusters are selected (using lasso selection), then the system makes a new cluster from these lasso picked data samples. 
\vspace{0.5em}

\noindent \textbf{Sub Clustering (T2)}: Hovering over a cluster reveals a plus button. Users can click on it to open a subcluster panel on the Cluster View, which shows subgroups of the data items within the selected cluster. In addition, a bar chart shows the distribution of a chosen attribute. Alongside, text description highlights the attributes that were used to compute the sub-clusters. Given that the users are not experts in data science, we do not present the quality metrics (e.g., silhouette scores, homogeneity score, etc.) Instead, we describe cluster models by showing thumbnail previews of clustering results with text descriptions as Figure~\ref{fig:comparison} shows.
\vspace{0.5em}

\noindent\textbf{Delete data items or Clusters (T3)}: Our discussion with biologists revealed that they sometimes need to `exclude' data items or clusters from their analysis while testing a hypothesis. 
Thus, we initially implemented the `delete' feature by enabling users to select a subset of items or clusters from the main view and click on the delete icon. 
However, when we showed it to the biologists, they had trouble due to inconsistencies between the button-based interaction and other demonstration-based interaction. 

The latest version of Geono-Cluster enables users to drag and drop a selected cluster on the delete icon shown on the top-left of the interface to demonstrate their interest in moving the selected cluster out of the layout. Similarly, they can drag-and-drop individual data items to demonstrate their interests in removing them from the cluster assignment. 
\vspace{0.5em}

\noindent\textbf{Creating Customized Clusters}: 
Users can select a subset of data items by clicking on the rows shown on the Table View (each row represents a data item). 
After selecting a subset of rows, users can drag-and-drop them on the Cluster View to demonstrate their interest in creating a clustering, in which all the selected data items fall in the same cluster. 
Users can iteratively repeat the process, and each drag-and-drop operation forms a new cluster in the Cluster View. 
Participants liked this idea as they found the design and the workflow of this interaction consistent with other interactions.

\subsection{Computational Techniques}
\label{Sec:comp_tech}

This section describes the underlying computational techniques which enable Geono-Cluster to recommend cluster models by incrementally steering (multiple cluster models) them to adhere to demonstrated user preferences.
Our cluster model recommendation process includes the human in the loop. On a high level, the user shows their intentions on a cluster layout. Based on the operations, Geono-Cluster models multiple cluster algorithms and finds top $k$ closest cluster models to the users' intention. Then, the user can refine the results through a series of customizations (instrumented through the interactions described above). In response, Geono-Cluster automatically finds close variants of cluster models and updates the recommendations in the Recommendation Panel. In summary, the system finds a set of cluster models with a distance function that reflects user-demonstrated cluster assignments.
\vspace{0.5em}

\noindent\textbf{Multiple clustering models:} 
The clustering task begins when the user requests a new cluster layout (when they press the cluster button in the interface). In response, Geono-Cluster generates multiple clustering models $M$.
Each cluster model $M_{i}$ in $M$ ($M_{1}$,$M_{2}$,$M_{3}$,$M_{4}$, ... $M_{T}$) is defined by a careful combination of a learning algorithm $\omega_{i}$ and a set of $p$ hyperparameters $\phi$, defined as $\phi_{i1}$, $\phi_{i2}$, $\phi_{i3}$, $\phi_{i4}$, ... $\phi_{ip}$. Applied clustering algorithms include K-Means, DBScan, Agglomerative Clustering, and Spectral Clustering.
Each algorithm has its hyperparameters.
For example, K-Means is a learning algorithm with ''k'' and the ''max-iteration'' value as an input hyperparameter.
Each model $M_{i}$ in $M$ is assigned a metric score $S_{i}$, which defines the quality of the clustering output (a higher $S_{i}$ means a better cluster definition). 
Geono-Cluster uses Scikit-Learn's ML package to construct and evaluate the cluster models using various quality metrics (e.g., Silhouette Coefficient, Davies-Bouldin index \cite{sklearnMetrics}, etc.)
\vspace{0.5em}

\noindent\textbf{Recommendation Technique:} 
Geono-Cluster ranks the models in $M$ by their scores $S$ and visualizes the best clustering layout in the Cluster View.
Further, the system allows the user to inspect top $f$ best cluster models from the ranked set of models $M$, through the Recommendation Panel (see Figure~\ref{fig:Geono-Cluster}-a).
If a user makes any customization to the shown cluster model $M_{c}$ (e.g., merge or split clusters),  the system automatically updates the recommendations by computing a new set of $M$ cluster models, except the model $M_{c}$, which is currently shown in the Cluster View. 
In doing so, the custom cluster assignments (i.e., $C = C_{1}, C_{2}, C_{3}, ... C_{m}$) for $m$ clusters in $M_{c}$) are used as ground truth labels by the system. 
On computing the new cluster models $M$, the system computes the performance of each model $M_{i}$ in $M$ by matching the predicted cluster assignments $A$ with the ground truth labels $C$. 
Using $A$ and $C$, the system updates $S$ defined by clustering metrics that are based on ground truth labels. 
We tested scores such as Fowlkes-Mallows score, Mutual Information based score, Adjusted Rand index, and Homogeneity \cite{sklearnMetrics}. For example, the Homogeneity score (which we used for the user study) shows how each cluster only contains items from a single class and how many members of a given class are assigned to the same cluster. 
Based on the updated score $S$, the system ranks the $M$ models. Next it visualizes the best model in $M$ in the Cluster View and shows thumbnail previews of the top $f$ models in the Recommendation Panel (\textbf{G2}).

Geono-Cluster's model recommendation finds the closest fitting cluster assignments, whenever the user customizes the current cluster layout in the Cluster View. However, there can be scenarios that no cluster recommendation matches the user's intended changes. This may occur when users seek clustering results, which are mathematically infeasible. There could be various reasons for it, such as, users may have a different understanding of the data than what the data actually contains, or the data may have noise, etc. In such cases, users may need to be educated to understand the reasons for a different clustering result, which we plan to integrate in the workflow in the future. 
Currently, in such cases, Geono-Cluster still responds with the nearest best clustering output, though it may not resemble the layout shown by the user. 
\vspace{0.5em}

\noindent\textbf{User driven feature selection:} A cluster model $M_{i}$ is driven by a set of features $F$ =  $f_{i1}$, $f_{i2}$,$f_{i3}$,$f_{i4}$ ....... $f_{ik}$ as input to compute the distance function which assigns a set of data items $D$ to individual clusters $C$. In Geono-Cluster, the set of features $F$ is either computed using feature selection methods e.g., ''select K Best'' \cite{scikitPedr}, and ''PCA'' \cite{MalhiPCA}, or can be retrieved from users if they specify a set of features and their relative weights (from the Attribute Panel supporting the task \textbf{T3}).
When users specify a set of $k$ features $F_{u}$ =  $f_{i1}$, $f_{i2}$,$f_{i3}$,$f_{i4}$ ... $f_{ik}$ with respective weights for each feature ($W_{u}$ =  $w_{i1}$, $w_{i2}$,$w_{i3}$,$w_{i4}$ ... $w_{ik}$, the system updates the distance function in the clustering algorithm.
The distance function is represented as $\sum_{i=1}^{k}\sum_{j=1}^{n}\left \| x_{i}^{(j)} * w_{i}^{(j)} - c_{j}^{} \right \| ^{2}$, where $c_{j}$, is the $j$th cluster centroid and $w_{i}^{(j)}$ is the user assigned feature weight.
\vspace{0.5em}

\noindent\textbf{Sub Clustering:} When triggered by users, the system builds a sub-cluster model $M_{si}$, for data instances $E$, member of a selected cluster $C_{i}$. Unlike the set of main cluster models $M$, only a single sub-cluster model is generated per cluster (\textbf{T2}). 
However, clicking on the ''add subcluster'' button again for the same selected cluster $C_{i}$, the system recomputes the sub-cluster model $M_{si}$, by choosing a new set of a learning algorithm $\omega$ and hyperparameters $\phi$; e.g., it picks a new ''k'' on the ''K-Means" cluster model. 
This technique allows users to rapidly browse a large set of sub-cluster models. 
\vspace{0.5em}

\noindent\textbf{Similar item selection:} Users click on a cell ($q_{j}$) of a quantitative attribute on the Table View to select a value $v_{j}$ of the data item $d_{i}$. Geono-Cluster finds a set of $r$ data instances,  $U$ = $d_{a}$,$d_{b}$,$d_{c}$ ... $d_{r}$, each of whose value $v_{j}$ falls within a threshold range, say [$+eps, -eps$]. The parameter $eps$ is set for each quantitative attribute $Q$ by heuristics and can be adjusted. This technique allows users to pick data instances which are similar, based on the selected quantitative attribute $q_{j}$. Further, users can select another quantitative attribute cell $q_{k}$. Next, from the set of selected data instances $U$, the system finds all instances $V$ which fall within a threshold range of the value selected for attribute $q_{k}$. Here the size of $V$ is less than that of $U$. This technique allow users to filter and select a subset of data instances $V$ from the Table View.
For categorical features $X$, Geono-Cluster performs exact feature value matching instead of matching data items based on a predefined range ([$+eps, -eps$]).

Users can drag-drop these $V$ data items to the Cluster View as a single cluster ($C = C_{1}$). They can continue selecting another set of data items, then add them to the cluster view as a new cluster ($C = C_{1}, C_{2}$). 
Users complete the data exploration or they can request the system to find a model $M_{i}$ iteratively (\textbf{T3}).

\noindent \textbf{Scalability:}
Unsupervised learning is expensive. As the number of data items increases, the cost of cluster assignments also grows higher. Geono-Cluster can run cluster computations for approximately $3000$ data items without major delays. For the scope of our study, the number of data items seem practical. 

\section{Evaluation}
To evaluate Geono-Cluster, we performed a qualitative assessment with six biologists to collect subjective feedback and observational data. Our study had two main goals: (1) collect qualitative feedback on Geono-Cluster's features and design, and (2) observe how experts perform visual clustering analysis using Geono-Cluster. In particular, our study indicates how Geono-Cluster helps domain experts gain insights into data by interactively building clusters. Thus, we also report some interesting findings that our participants generated throughout the sessions.

\subsection{Participants and Setting}
We recruited $6$ biologists ($2$ female, $4$ male), all with graduate degrees related to Biology, Bio-Statistics or Bio-Informatics. They had $1-2$ years of experiences working with \textit{Gene} related datasets. They had not participated in our preliminary evaluation of Geono-Cluster and were also not involved in the design of Geono-Cluster. All participants were familiar with the concept of data clustering and had previous experience with data grouping with at least one data analysis tool (e.g., SAS, R, etc.). Further, as they had previously worked with \textit{GWAS} catalogue data, they were familiar with all the data attributes in the dataset. During the entire study participants used a computer with 17-inch screen and used a mouse to interact with the system. The study took approximately $50$ minutes and we rewarded each participant with a \$ $20$ gift card.

\subsection{Procedure}
\noindent\textbf{Introduction and Training:} Participants were briefed about the purpose of the study and their rights. After filling out the study consent form and a questionnaire on demographics, we asked participants to watch a tutorial video of Geono-Cluster. The video walked the participants through different features and interactions provided by the tool. After watching the video, we asked participants to work with the tool for $10$ minutes. In addition, we encouraged the participants to ask as many questions as they want during this stage. 
\vspace{0.5em}

\noindent\textbf{Main Study:}. The participants were asked to explore the \textit{GWAS} Cataloge~\cite{GWAS_catalog} data that includes published \textit{SNPs} and association studies. In particular, we asked the participants to imagine their colleagues asked them to analyze the dataset using the visualization tool for $30$ minutes and report their findings. Participants were instructed to verbalize analytical questions they have about the data, the tasks they perform to answer those questions, and their answers to those questions in a think-aloud manner. In addition, we instructed them to come up with data-driven findings rather than making preconceived assumptions about the data. 
The interviewer played a role of 'active listener' during the study. He facilitated participants' verbal reports by asking questions like ``what are you trying to do?'', ``what are your thoughts now?'', ``what do you think about current groupings?''. We tried to avoid interrupting the participants as much as possible during their data exploration process. However, we sometimes reminded that this is a think-aloud study and they need to verbalize their thoughts. 
\vspace{0.5em}

\noindent\textbf{Follow-up Interview:} After each participant complete the task, the experimenters asked the following questions:

\begin{enumerate}[leftmargin=4mm,itemsep=0cm]
\item Tell me about your experience with this tool.

\item What did you like or dislike about this tool?

\item Did this tool help you in your data exploration? If so how?

\item Is there anything else that you want to add?

\item What are the major obstacles/roadblocks while using the tool to solve your problems? How did you go around (resolve) the issue? Do you have ideas to improve the tool?

\end{enumerate}
\vspace{0.5em}

\noindent\textbf{Wrap-up}. The experimenter thanked the participants who received a \$20 value Amazon gift card. 

\subsection{Data Collection and Method of Analysis}
We screen- and audio-recorded the whole study. During the main study, the experimenter took notes while participants interact with the system. We also collected feedback from a semi-structured interview with open-ended questions at the end of the study. 

We analyzed around 300 minutes of screen-capture videos from six participants. First, one of the authors transcribed the audio recording of the study. Then, two coders (first and second authors) read the transcribed data (including the think-aloud sessions and the interview responses) to parse a set of meaningful text snippets. After reading the data, each of the coders independently assigned codes (a word or phrase) to best describe the text snippets. Finally we consolidated the codes from the two authors by focusing on the aspects of the responses which highlighted positive or negative feedback with respect to \textit{usability of the system}, \textit{easy of use}, \textit{learning curve}, \textit{future feature requests} or \textit{strategies pertaining to exploratory data analysis using clustering models}. In the following section, we use \textbf{P1} to \textbf{P6} to respectively denote the participants one to six who participated in the evaluation. 

\subsection{Results and Feedback}
Overall, all participants found Geono-Cluster easy to use and effective in performing cluster analysis tasks. However, a few of them experienced difficulties in interpreting the recommendations made by the system. Below, we categorize and discuss the findings of our qualitative study in more details.
\vspace{0.5em}

\noindent \textbf{System usability}: 
All participants found Geono-Cluster's workflow easy to use, intuitive, and engaging. P2 remarked \textit{``I can keep trying new ideas to quickly test different ways to cluster this data.''} P4 said \textit{``It's so easy to use, I can quickly iterate and learn about the data much faster, than using packages in R to cluster data.''} Further, many other participants found visualization to be a very good medium to learn about the data by exploring different clustering results. P5 said \textit{``I never knew that I can use visual methods to explore clustering result. Currently I use R to cluster my data, then export a CSV file to my team-mates.''}
\vspace{0.5em}

\noindent \textbf{Consistency with user mental model}: 
Participants found the design and workflow of Geono-Cluster consistent with their mental model and expectations. In particular, participants found that it is intuitive to visually demonstrate tasks such as creating, merging, and splitting clusters by demonstration. For example, P3 mentioned: \textit{``it feels intuitive to merge clusters by dragging and dropping one cluster over another one. [...] this is what I would expect to happen.''} P5 stated: \textit{``I liked the idea of creating a cluster of items by moving the data items from this table to the empty space [dragging the data items from the Table View and dropping them on the Cluster View to create a cluster].''} Further P2 added: \textit{``Compared to programming, using this kind of tool is more straight forward and faster.''} Consistency and natural mapping between user's intent and the actions required for performing the intent is important in designing new interactions~\cite{norman1993things, perin:2014:direct}. Consistency also improves learnability and ease of use of an interaction~\cite{perin:2014:direct}. 
\vspace{0.5em}

\noindent \textbf{Perceived control over data analysis process}: 
While using Geono-Cluster, P1, P4, and P5 commented on their level of control over the data analysis that resulted from their freedom in interacting with visualizations instead of going through layers of menu items. For example, P1 mentioned: \textit{``This is great because I can construct my own cluster and tell the system how I want my clustering outcome looks like.''} P4 stated: \textit{``It is a powerful idea to enable analysts to use their knowledge about the data items to interactively create clusters [visually demonstrate their expected clustering outcome]. I specifically like how this allows merging and splitting clusters.''}
The level of interaction directness~\cite{beaudouin2000instrumental} with the visual representation contributes towards increasing the perceived control of the participants over the data analysis process. 
\vspace{0.5em}

\noindent \textbf{Difficulty in splitting a cluster}: 
Overall, participants found the lasso interaction intuitive and easy to use. However, with lasso selection participants were not very exact about the data items that they wanted to select. For example, after selecting a subset of data items, P3 noted: \textit{``It is hard to be exact with this selection. I don't want this specific point to be selected.''} In such cases, participants had to either deselect the items that were selected incorrectly by clicking on them or try to lasso select again. Going forward, we envision designing advanced interaction techniques for easier selection of data items that are located in a close distance from one another.
\vspace{0.5em}

\noindent \textbf{Interpretability of recommendations}: 
While using Geono-Cluster, participants were sometimes unsure why the system suggested specific recommendations. While designing Geono-Cluster, we decided to show each recommendation as a thumbnail on the Recommendation Panel. 
Each recommendation uses the natural language to explain the most representative features used for the clustering. 
In addition, each recommendations shows the resulting outcome of the recommended cluster. 

Although some participants liked how the recommendations were presented, two participants could not immediately understand why specific recommendations are suggested. For example, P2 mentioned:\textit{ ``I understand what each cluster represents which is good, but I am not sure why these recommendations.''} and P3 stated: \textit{``I am curious how these recommendations are added.''}. Going forward, we suggest systems to explore design alternatives
to explain the reasoning behind recommendations.
In situations when the system does not find any cluster recommendations that matches user's demonstrated changes, Geono-Cluster shows the nearest best clustering layout. 
In such scenarios, users may be surprised to see the abrupt or strikingly different recommendations.
In the future, we are thinking of explicitly communicating this conflict in textual description.
At the same time, we want to introduce a more variety of models so that the system can perform deeper search to find desirable results.
\vspace{0.5em}

\noindent \textbf{Custom labeling and annotation feature}:
Two participants found that with growing number of clusters, it became hard for them to remember what each cluster represents and how they came to the cluster. Thus, they suggested adding a feature that enables them to annotate and label the clusters and to record the operations performed on clusters beforehand. 
For example, P6 said \textit{``it will be nice to know what each cluster represents, meaning every cluster should have an annotation, or users can add custom annotations. May be label the cluster by the most prominent feature of the cluster.''} P3 mentioned: \textit{``Is there a way to label each cluster?''}


\section{Observations}
Our user study reveals that participants usually began exploring the data by framing a hypothesis, asking the questions they want to know, and then performing a set of tasks (as described in section \ref{sec:tasks}) through Geono-Cluster's interface to find the answers. 
Interestingly, we observed that participants often took two different approaches to perform visual data clustering: \textbf{Top-down} and \textbf{Bottom-up}. Below, we describe each approach in more details.


\subsection{Top-down Visual Data Clustering Approach}
P1 started his data analysis process by asking \textit{``How does the gene samples differ in disease risk factor by regions and chromosome factors?''} To that end, P1 clustered data items by selecting a set of features from the Attribute Panel and then pressed the \textit{Cluster} button. Next, he checked the recommended cluster layouts from the Recommendation Panel to explore other clustering results based on another set of features. In response, he updated the list of features to cluster the data by and triggered Geono-Cluster to generate a new cluster layout. P4 also followed the same approach; however, he did not have any question to begin with. 
He initialized the process by pressing the cluster button to start with an initial clustering. Next, he hovered over data items in each cluster to familiarize himself with the data items and find similarity or dissimilarity. He also checked the Table View to compare different data items from various clusters. If the clusters did not match his mental model, he would adjust the features from the Attribute panel. He would then preview the recommended clustering options to further explore a wide range of cluster outputs. This process continued until he was satisfied with the clusters and had a better sense of the data.


A main point here is that in the top-down approach participants mostly avoided interaction at the data item level, but instead they dealt with the full range of features from the Attribute Panel. P1 also verified this point by saying: \textit{``I relied on cluster button to cluster the data, as I do not specifically know much about the data items, so did not use the table's drag-drop feature. Similarly, I did not customize the clusters by using lasso or drag-drop feature initially. I rather re-computed the clusters based on a new set of features that I specify.''} However, P1 later confirmed that over iterations when he was more confident about the data, he started using the split and merge operations to customize shown clusters.

\subsection{Bottom-up Visual Data Clustering Approach}
Remaining participants (P2, P3, P5, and P6) followed the Bottom-up approach, in which they mainly relied on interaction at the data item level. They first created a customized cluster by dragging data items from the Table View and dropping them on the Main view as opposed to relying on the cluster button. These participants often interacted with data items to demonstrate their expected outcome. 

P2 started her clustering analaysis by asking \textit{``How does the gene samples derived from humans/monkeys (ANC) vary from gene samples derived from mixing humans and monkeys (DER) with respect to various diseases?''}. To answer the question, P2 placed all the \textit{ANC} gene samples into one cluster and a few \textit{DER} gene samples into another cluster from the Table View. P2 remarked: \textit{``my strategy is to select a set of data points [items] based on the gene's ancestry, then drag-drop to create a cluster''}.
P2 then previewed the recommendations to explore other options to cluster the data based on his specification of clusters. In this process, P2 did merge/split clusters to test different ideas to cluster the data using the lasso-selection and the cluster drag-drop feature. P2 said: \textit{``I also rely on the lasso tool to define other clusters from this, if the cluster appears too big''}. Using Geono-Cluster, many participants were able to customise clusters in this fashion to find interesting insights from the data, that they found needed further analytical investigations/research with their peers or mentors.
For example, one participant was able to find a significant difference in \textit{Average-risk-allele-frequency} between two sets of clusters by iteratively following this bottom-up visual clustering approach.

P6 also followed the same approach. P6 mentioned: \textit{``I want to know if the gene with chromosome factor higher than 6 sampled from America, have higher cancer risk factor? 
To seek an answer, I find the Table View's data item selection feature quite useful, as I can define my own clusters based on chromosome value or the region the gene was sampled from.''}
We noticed that when P6 explored the initial set of cluster layouts, he paid attention to the suggested features (in the recommended cluster layouts) to understand how the cluster is defined.
In some cases, P6 did not agree with the recommendations or the features that were used to derive the results. To provide his feedback for updated results, he customized the best-perceived cluster layout by splitting the existing clusters using the lasso tool and merging smaller clusters into one. P6 added: \textit{I am using the lasso feature to take out all the data items which have chromosome value less than 6. Also, the smaller clusters with 5 or fewer data samples are confusing, so I merge them into one}. P6 continued customizing cluster definitions and triggering Geono-Cluster to seek better cluster layouts until the output matched his mental model of the data.

\section{Discussion}
In this section, we discuss lessons learned from our study. We also discuss limitations of our approach for future studies. \vspace{0.5em}

\noindent\textbf{Model Feedback and Interpretation:}
Periodic discussion and informal inputs from the biologists clarified that model interpretation and feedback (to the model) is of critical value to them. For example, when Geono-Cluster shows a set of clustering recommendations, users may need to know how they differ from each other, or what logic was implanted to define the displayed clusters. 
There are many ways to explain this to the user; however, we only selected methods which does not require any technical expertise from the user. 
Our final design explains a cluster by using a natural language-based approach to communicate the features that were used to compute the clustering distance function. In particular, we avoid showing technical information such as silhouette coefficient or exact feature weights to provide a high-level model explanation that does not overwhelm the users with a bag of information that might not be easy to interpret. Our qualitative feedback hints that our approach made Geono-Cluster not only easy to use but also an engaging tool to continue data exploration by rapidly testing different ideas to cluster the data. 
\vspace{0.5em}


\noindent\textbf{Cluster Model Comparison:}
While representing multiple clustering results show different ways to partition the data, model comparison to understand trade-offs between these clustering options is critical. However, in our current prototype we do not support explicit cluster model comparison. For example, users cannot perform a pairwise comparison of two cluster models side by side~\cite{Gleicher_comparison}, or they cannot select a few chosen cluster models to see the results in a way which facilitates direct comparison. 
Based on our interviews with the biologists, comparing cluster models was not posed as a requirement to us. Therefore, we deliberately did not include cluster model comparison as one of the design goals of the system. However, as visual analytics researchers, we understand that being able to compare multiple cluster models, may positively aid model selection and enhance the tools use case.
\vspace{0.5em}

\noindent\textbf{Limited Model Explanation:}
Geono-Cluster explains a cluster model by highlighting the top $k$ features that were used to compute the underlying distance function using a natural language expression. Though the simplicity of the explanation is helpful for non-experts, in certain cases this may pose as a very limited explanation of a clustering model. For example, two cluster models may be based on the same set of features, but the defined clusters are strikingly different. In this case, users may get confused to interpret the difference between these models. We reckon this as a limitation of our tool which we intend to study in future work. 
\vspace{0.5em}

\noindent\textbf{Scalability:}
The current interface and the supported interactions (i.e., split and merge technique) is tested with $3000$ (approximately) data items. However, we understand that as the size of the data grows, the interaction techniques such as drag-and-drop interaction and lasso-selection tool may be less responsive. In the user study, P6 noted that the lasso-selection was less effective for large clusters when the data items became too small to select or notice (often partially obscured by neighboring data items). We envision multiple ways to enhance scalability in such demonstration-based systems. For instance, we could aggregate data items as defined by the user, e.g., group all data items with \textit{Chromosome} value greater than $7$.
\vspace{0.5em}

\noindent\textbf{Annotation Feature:} While working with Geono-Cluster, some participants expressed their interest in having a feature that enables them to annotate different clusters based on their prior knowledge or insights they gained during data analysis. For instance, one participant mentioned her interest in annotating clusters based on prominent features to help understand cluster compositions easily. Going forward, we consider this feature for the next iteration fo the tool development to further enhance data exploration and pattern finding using clustering methods.

\section{Conclusion}
In this paper, we introduce Geono-Cluster that is designed to help biologists visually cluster their data for exploratory analysis. The proposed technique leverage the domain knowledge of the users by allowing a demonstration based interaction methodology, which recommends multiple cluster models according to users' intent. 
Based on collaborative studies with biologists, we built a set of task requirements and design guidelines for our prototype. 
The technique shown exemplifies a model of interaction which allows non-experts in data science interactively construct clustering models by specifying their preferences. 
This spares them the burden of going through layers of menus and control panels to transform their expectations to outputs or to comprehend complex model parameters or metrics to find the right clustering model.
Our study provides valuable lessons for researchers who design visual clustering tools for biologists.
\ifCLASSOPTIONcaptionsoff
  \newpage
\fi



\bibliographystyle{IEEEtran}
\bibliography{IEEEabrv,template}
%




%

\begin{IEEEbiographynophoto}{Bahador Saket}
is currently a fifth year PhD student at the Georgia Institute of Technology. His research combines methods from data visualization and human-computer interaction (HCI) to develop and evaluate natural interaction techniques for visual data analysis, and apply these to interactive systems in different critical domains. 
\end{IEEEbiographynophoto}

\begin{IEEEbiographynophoto}{Subhajit Das}
is a Ph.D. Computer Science student at the School of Interactive Computing, Georgia Institute of Technology, USA. His research interests include interactive machine learning, model optimization/selection, and designing human-in-the-loop based visual analytic systems. 
\end{IEEEbiographynophoto}

\begin{IEEEbiographynophoto}{Bum Chul Kwon}
is Research Staff Member at IBM Research. His research area includes visual analytics, data visualization, human-computer interaction, healthcare, and machine learning. Prior to joing IBM Research, he worked as postdoctoral researcher at University of Konstanz, Germany. He received his M.S. and Ph.D. from Purdue University in 2010 and 2013, respectively. He received his B.S. in Systems Engineering from University of Virginia in 2008.
\end{IEEEbiographynophoto}



\begin{IEEEbiographynophoto}{Alex Endert}
is an Assistant Professor in the School of Interactive Computing at Georgia Tech. He directs the Visual Analytics Lab, where him and his students explore novel user interaction techniques for visual analytics. His lab often applies these fundamental advances to domains including text analysis, intelligence analysis, cyber security, decision making, and others. He received his Ph.D. in Computer Science at Virginia Tech in 2012.
\end{IEEEbiographynophoto}




\end{document}